\documentclass{PoS}
\usepackage{amssymb} 
\usepackage{amsmath}
\usepackage{mathptmx}
\usepackage{calrsfs}
\usepackage[caption=false]{subfig} 
\newcommand{\nn}{\nonumber}
\DeclareMathAlphabet{\mathcal}{OMS}{zplm}{m}{n}
\newcommand{\Q}{{\cal Q}}

\newcommand{\ce}[1]{Eq.~(\ref{#1})}
\newcommand{\cf}[1]{{Fig.~\ref{#1}}}

\renewcommand{\d}{\mathrm{d}}
\newcommand{\sT}{{\scriptscriptstyle T}}

\newcommand{\CS}{{\rm CS}}

\newcommand{\mc}[1]{\mathcal{#1}}
\newcommand{\kT}{\pmb{k}_T}
\newcommand{\koneT}{\pmb{k}_{1T}}
\newcommand{\ktwoT}{\pmb{k}_{2T}}
\newcommand{\qT}{{\pmb{P_{f}}_T}}

\newcommand{\eqs}[1]{\begin{equation} \begin{split} #1\end{split} \end{equation} }
\def\ie{{\it i.e.}}
\def\eg{{\it e.g.}}

\let\OLDthebibliography\thebibliography
\renewcommand\thebibliography[1]{
  \OLDthebibliography{#1}
  \setlength{\parskip}{0pt}
  \setlength{\itemsep}{-3pt}
\footnotesize
}

\title{Probing the gluon TMDs with quarkonia}

\ShortTitle{Probing the gluon TMDs with quarkonia}

\author{\speaker{Jean-Philippe Lansberg}\\
        IPN Orsay, Paris Sud U., CNRS/IN2P3, Universit\'e Paris-Saclay, F-91406 Orsay, France\\
        E-mail: \email{Jean-Philippe.Lansberg@in2p3.fr}}

\author{Cristian Pisano\\Dipartimento di Fisica, Universit\`a di Cagliari, and INFN, Sezione di Cagliari
    Cittadella Universitaria, I-09042 Monserrato (CA), Italy}

\author{Florent Scarpa\\
       IPN Orsay, Paris Sud U., CNRS/IN2P3, Universit\'e Paris-Saclay, F-91406 Orsay, France \&
Van Swinderen Institute for Particle Physics and Gravity,
University of Groningen, Nijenborgh 4, 9747 AG Groningen, The Netherlands}
\author{Marc Schlegel\\ Department of Physics, New Mexico State University, Las Cruces, NM 88003, USA}

\abstract{We briefly review how quarkonium hadroproduction can be used to access the polarised and unpolarised gluon TMDs.}

\FullConference{XXVI International Workshop on Deep-Inelastic Scattering and Related Subjects (DIS2018)\\
		16-20 April 2018\\
		Kobe, Japan}

\begin{document}

\section{Introduction}\vspace*{-0.4cm}

Transverse-Momentum-Dependent (TMD) factorisation~\cite{Collins:2011zzd,Aybat:2011zv,GarciaEchevarria:2011rb,Angeles-Martinez:2015sea} 
allows one to study the impact of the polarisation of partons with nonzero transverse momentum --even inside unpolarised hadrons--
via the appearance of azimuthal modulations. At hadron colliders, these modulations mostly come from gluons and these 
new phenomena are encapsulated in the distribution $h_1^{\perp\,g}(x,\kT^2)$ 
of linearly-polarised gluons~\cite{Mulders:2000sh}. In practice, one expects $\cos 2\phi$ ($\cos 4\phi$) modulations in the yield of
two-particle final states. These follow from single (double) gluon-helicity flips in gluon-fusion processes. The correlation between the polarisation of these gluons and their transverse momentum
 can also alter the transverse-momentum spectrum of the produced system. One example is that of the Brout-Englert-Higgs $H^0$ boson~\cite{Boer:2011kf,Boer:2013fca} produced after double gluon-helicity flips. 

We review here how some processes involving quarkonium production can help experimentally determine these poorly known gluon distributions.
First, we start with the production of a pseudoscalar quarkonium which is very similar to the $H^0$ case. Second, we discuss 
the production of a $J/\psi$ or $\Upsilon$  in association with a photon or a $Z^0$ boson. Third, we report on the unique case of 
di-$J/\psi$ production which is expected to exhibit the largest possible $\cos 4\phi$ modulations and for which data exist allowing for the 
first extraction of the unpolarised gluon distributon $f_1^{g}(x,\kT^2)$ using recent LHCb data.

\section{TMD factorisation and quarkonium production}\vspace*{-0.2cm}

As clear from its name, TMD factorisation extends collinear factorisation by accounting
for the parton transverse momenta, generally denoted $\kT$. In the case of quarkonium-production processes, it is applicable either when 
a single quarkonium  $\Q$ is produced with a transverse momentum, $P_{fT}$, typically smaller than half of its mass or when a quarkonium is produced in a set of two particles whose transverse momentum, $P_{fT}$, is also smaller than half of its invariant mass, $M_f$. In both cases, the observed final state should be colourless (see \eg~\cite{Boer:2016bfj}). This imposes the quarkonia to be produced by Colour Singlet (CS) transitions (or equally speaking according to the CS Model (CSM)) and excludes the production in association with light hadrons, charm or beauty hadrons. 
 
Under the TMD factorisation, the cross section for any gluon-fusion process 
can be expressed --up to corrections suppressed by powers of the  observed-system transverse momentum--
as a contraction and a convolution
of a partonic short-distance contribution, $\cal M_{\mu\rho}$, with two gluon TMD correlators $\Phi_g$
evaluated at $(x_1,\koneT)$ and $(x_2,\ktwoT)$.
$\cal M^{\mu\rho}$ is simply calculated in perturbative QCD through a series expansion 
in $\alpha_s$~\cite{Ma:2012hh} using Feynman graphs. Overall, we have
\begin{multline}\label{eq:factformula}
\d\sigma = \frac{(2\pi)^4}{8 s^2}\!
  \int\!\! \d^{2}\koneT \d^{2}\ktwoT
  \delta^{2}(\koneT + \ktwoT - \qT)
  \mc{M}_{\mu\rho}
  \left(\mc{M}_{\nu\sigma}\right)^*\\
\times   \Phi_g^{\mu\nu}(x_1,\koneT)\,
  \Phi_g^{\rho\sigma}(x_2,\ktwoT)\, \d\mc{R}\,,
\end{multline}
where $s = (P_1 + P_2)^2$ is the hadronic centre-of-mass system (c.m.s.) energy squared, $P_f$ is the momentum of the observed final state,
and where the phase-space element of the outgoing particles is denoted by $\d\mc{R}$. In addition, the gluon four-momenta $k_i$ are decomposed according to $k = xP + k_{\sT} + k^-n$ [$n$ refers to a light-like vector ($n^2=0$) satisfying $n\cdot P \neq 0$], $k_{\sT}^2 =-\kT^2$ and  $g^{\mu\nu}_{\sT} = g^{\mu\nu} - (P^{\mu} n^{\nu}+ P^\nu n^\mu)/P{\cdot} n$.

In the case of unpolarised protons, the correlator can be parametrised~\cite{Mulders:2000sh,Meissner:2007rx,Boer:2016xqr} in terms of the two aforementionned and independent TMDs,	the unpolarised distribution $f_1^g(x,\kT^2)$ and the distribution of linearly polarised gluons $h_1^{\perp\,g}(x,\kT^2)$,\footnote{
For gluon TMDs free from rapidity divergences from gauge links with paths (partly) along the light front $\xi\cdot n =0 $, a {\it soft factor}~\cite{Collins:2011zzd,Echevarria:2012js,Echevarria:2015uaa} is required. It  does not play a role here and we therefore consider it as implicit.}
	      	      \begin{align}
\Phi_g^{\mu\nu}(x,\kT) = 	-\frac{1}{2x} \,\bigg \{g_\sT^{\mu\nu} f_1^g
	-\bigg(\frac{k_\sT^\mu k_\sT^\nu}{M_p^2}\,
	{+}\,g_\sT^{\mu\nu}\frac{\kT^2}{2M_p^2}\bigg)\,
	h_1^{\perp\,g} \bigg \} \,,
\end{align}

Following~\cite{Lansberg:2017tlc}, the structure of the TMD differential cross section 
for an observed system of 2 colourless particles reads
\begin{align}\label{eq:crosssection-2-particles} 
&\frac{\d\sigma}{\d M_{f} \d Y_{f} \d^2 \qT \d \Omega} 
  =
{\cal J} \times
\Big \{
  F_1\, \mc{C} \Big[f_1^gf_1^g\Big]  +
  F_2\, \mc{C} \Big[w_2h_1^{\perp g}h_1^{\perp g}\Big]+  \nn \\&
\cos2\phi_{\CS} \Big(F_3 \mc{C} \Big[w_3 f_1^g h_1^{\perp g}\Big]  + F'_3 \mc{C} \Big[w'_3 h_1^{\perp g} f_1^g \Big]\Big)  + \cos 4\phi_{\CS}F_4 \mc{C}\! \left[w_4 h_1^{\perp g}h_1^{\perp g}\right]\!\Big \}\,,
\end{align}
where $\d\Omega=\d\!\cos\theta_{\CS}\d\phi_{\CS}$, $\theta_{\CS}$ and $\phi_{\CS}$ are the Collins-Soper (CS) angles~\cite{Collins:1977iv} and
$Y_{f}$  is the pair rapidity -- $\qT$ and $Y_{f}$ are defined in the hadron c.m.s. In the {\CS} frame, the $\Q$ direction is along $\vec e=(\sin\theta_{\CS}\cos \phi_{\CS},
\sin\theta_{\CS}\sin \phi_{\CS}, \cos\theta_{\CS})$. The overall factor, ${\cal J}$, is 
specific to the mass of the final-state particles and the analysed differential cross sections, and
the hard factors $F_i$ do not depend on the rapidity of the pair nor on its transverse momentum.

For a single-particle (here quarkonium) production, 
\begin{align}\label{eq:crosssection-1-particle} 
\frac{\d\sigma}{\d Y \d^2 \qT} 
  = {\cal J}' \times \,\Big\{
  F_1\, \mc{C} \Big[f_1^gf_1^g\Big]  +  F_2\, \mc{C} \Big[w_2h_1^{\perp g}h_1^{\perp g}\Big]\Big\}
\end{align}

It is interesting to note that the TMDs appear in both above equations ~\ce{eq:crosssection-2-particles} 
and ~\ce{eq:crosssection-1-particle} in a factorised way from the hard-scattering coefficients through 
universal convolutions  which read 
\begin{multline}\label{eq:Cwfg}
\mathcal{C}[w\, f\, g] \equiv \int\!\! \d^{2}\koneT\!\! \int\!\! \d^{2}\ktwoT\,
  \delta^{2}(\koneT+\ktwoT-\qT) \, w(\koneT,\ktwoT)\, f(x_1,\koneT^{2})\, g(x_2,\ktwoT^{2}) \, ,
\end{multline}
where $w(\koneT,\ktwoT)$ are generic transverse weights and $x_{1,2} =  \exp[\pm Y_f]\, M_f/\sqrt{s}$. 
The $w(\koneT,\ktwoT)$ are identical
for all the gluon-induced processes with unpolarised protons and can be found in~\cite{Lansberg:2017tlc}. 
For any process, one can show that $F^{(')}_{2,3,4} \leq F_1$.

The azimuthal modulations can  be studied by evaluating   
[for $n=2,4$] 
weighted differential cross sections
normalised to the azimuthally independent term like 
\eqs{\displaystyle
\frac{\displaystyle \int \!\!d\phi_{\CS} \cos n\phi_{\CS}\,  \frac{d\sigma}{d M_{f} d Y_{f} d^2 \qT d \Omega}}
{\displaystyle\!\!\int \!\!d\phi_{\CS} \frac{d\sigma}{d M_{f} d Y_{f} d^2\qT d \Omega}}.}
One then gets that, in a single phase-space point, such $\cos 2 \phi_{\CS}$ modulations are proportional to 
$  F_3\, \mc{C} \Big[w_3 f_1^g h_1^{\perp g}\!\Big] + F'_3\, \mc{C} \Big[w'_3  h_1^{\perp g} f_1^g\!\Big]$
and the $\cos 4 \phi_{\CS}$ modulations  to 
$  F_4 \mc{C} \Big[w_4 h_1^{\perp g}h_1^{\perp g}\Big]$.

\section{Pseudoscalar-quarkonium production}\vspace*{-0.4cm}

The case of the pseudoscalar-quarkonium production was first studied under the TMD factorisation by Boer and Pisano in Ref.~\cite{Boer:2012bt}. It is very similar to that of $H^0$ boson and is particularly interesting since $F_2=-F_1$. From this, it follows that the  $\qT$ modulations
generated by $h_1^{\perp\,g}(x,\kT^2)$ are maximum. 
There are however some caveats. First, the domain where the TMD factorisation can be applied for $\eta_c$ production is admittedly very
small since $P_{\eta_c T}$ should remain below $1\sim 1.5$ GeV. As for now, $\eta_c$ production was only studied by LHCb~\cite{Aaij:2014bga} for $P_{\eta_cT}$ larger than 6 GeV. Studies at lower  $P_{\eta_c T}$ would require a specific effort on the triggers and on the background subtraction with however few hopes to reach $1\sim 1.5$ GeV. Studies in the fixed-target mode at $\sqrt{s_{NN}}=115$~GeV~\cite{Hadjidakis:2018ifr,Brodsky:2012vg,Lansberg:2012kf} with a smaller combinatorial background may be more promising. $\eta_c(2S)$ production 
may be another option~\cite{Lansberg:2017ozx}.
Slightly higher $P_{\eta_c T}$ may be dealt with by matching the TMD-factorised cross section with the collinearly factorised one with PDFs. However, this may preclude the extraction of experimental constraints on the TMDs~\cite{Echevarria:2018qyi}. Some of these caveats could be avoided by focusing one's effort on the $\eta_b$ case. However, it remains experimentally inaccessible at the LHC owing to its very small branchings to usable decay channels.

\section{$J/\psi+\gamma$ and $\Upsilon+\gamma$}\vspace*{-0.4cm}

As we have just seen, the restriction $\qT \lesssim M_f/2$ is extremely detrimental for the $2 \to 1$ case. To bypass this constraint, 
it is thus expedient to consider 2-particle final states where $M_f$ can be tuned to reach the optimal kinematical range for $\qT$
given the expected yields. Along these lines, we proposed~\cite{Dunnen:2014eta} to study the associated production of $J/\psi+\gamma$ and $\Upsilon+\gamma$ at the LHC whose QCD corrections have been studied in the collinear factorisation in~\cite{Li:2008ym,Lansberg:2009db,Li:2014ava}. 

If one is mainly interested in the extraction of $f_1^{g}$ via the $\qT$ dependence of the cross section, this process is interesting since
$F_2$ vanishes. As such, it is the complete opposite case compared to $\eta_c$ where the $\qT$  modulation from $h_1^{\perp g}$ are maximum. 
As what concerns the azimuthal modulations, $F_3$ is power suppressed in $M_f$ while $F_4$ scales like $F_1$ offering interesting opportunities to extract $h_1^{\perp g}(x,\kT^2)$ with existing LHC data by looking at $\cos 4\phi_{\CS}$ modulations. The ratio $F_4/F_1$ is however significantly smaller than unity and, in practice, the $\cos 2\phi_{\CS}$ modulations are not necessarily much smaller.
Studies in the fixed-target mode with a LHCb-like detector are feasible and may bring about constraints at $x$ as high as 0.5~\cite{Hadjidakis:2018ifr,Lansberg:2014myg}. 

Similarly, $J/\psi$ and $\Upsilon$ can be gluo-produced in association with an off-shell photon or a $Z$ boson. We studied this case in Ref.~\cite{Lansberg:2017tlc}. The rates are however probably too small~\cite{Gong:2012ah}, even at the LHC, for azimuthal-modulation studies. In addition, there may be a significant contamination by Double Parton Scatterings (DPSs)~\cite{Lansberg:2016rcx,Lansberg:2017chq} in the currently accessible region at the LHC~\cite{Aad:2014kba}.

\section{Vector-quarkonium-pair production}\vspace*{-0.4cm}

The third case we discuss here is that of the production of a pair of $J/\psi$ or $\Upsilon$.
As compared to the former processes, these reactions are extremely appealing for the following reasons:
\begin{itemize}\setlength{\itemsep}{-2mm}
\item the hard-scattering coefficients $F_i$ are extremely favourable to the observation of azimuthal modulations~\cite{Lansberg:2017dzg};
\item significant data sample have already been collected at the LHC and the Tevatron at different invariant masses~\cite{Aaij:2011yc,Abazov:2014qba,Khachatryan:2014iia,Aaboud:2016fzt,Aaij:2016bqq};
\item the DPS contamination is probably on the order of 10\%~\cite{Lansberg:2014swa,Aaboud:2016fzt}, where the azimuthal modulations are expected to be the largest;
\item the contamination from the non-TMD-factorising colour-octet transitions is also negligible --as expected from its $v^8$ suppression-- in this region~\cite{Ko:2010xy,Li:2013csa,Lansberg:2014swa}.
\end{itemize}	

As what concerns the hard-scattering coefficients, it is interesting to look at them both in  the 
large and small  $M_f=M_{\Q\Q}$  limits. Indeed, when $M_{\Q\Q}$ becomes much larger than the quarkonium mass, $M_{\Q}$, one finds~\cite{Lansberg:2017dzg} that, for $\cos\theta_{\CS} \to 0$ (\ie\ small $\Delta y$),    
\eqs{\label{eq:Fi_large_M}
F_4 \to  F_1, \quad
F_{2} \!\to\! \frac{
81 M_{\Q}^{4}\cos\theta_{\CS}^{2}}{2 M_{\Q\Q}^4} \times F_1,\quad
F_3 \!\to\!  \frac{- 
24 M_{\Q}^{2}\cos\theta_{\CS}^{2}}{M_{\Q\Q}^2} \times F_1.}

The result $F_4 \to  F_1$ in this limit is thus far unique among all the gluon-TMD-sensitive processes, making di-quarkonium production the most sensitive probe of $h_1^{\perp g}$.

\begin{figure}[hbt!]
\centering
\subfloat{\includegraphics[height=0.44\textwidth,angle=-90]{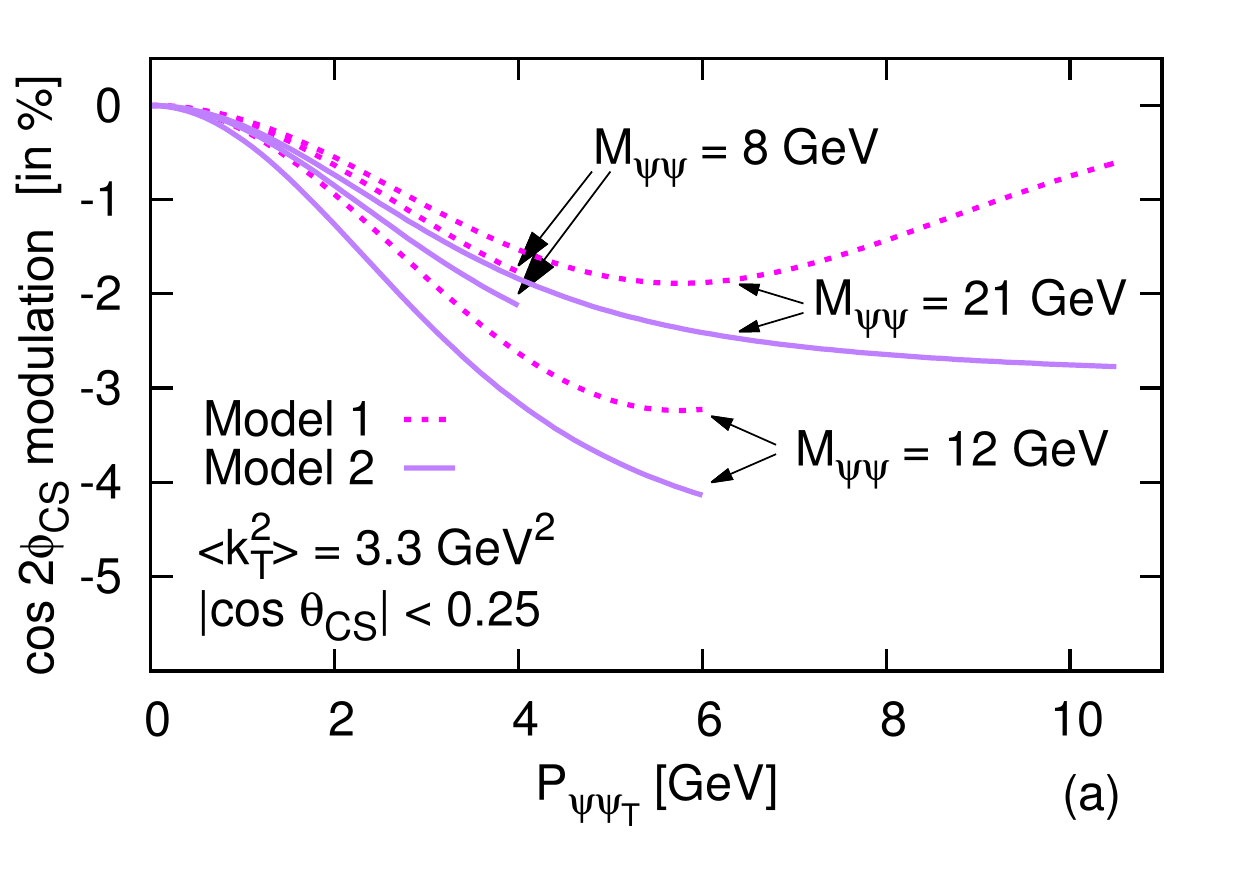}\label{fig:S2_costheta}}
\subfloat{\includegraphics[height=0.44\textwidth,angle=-90]{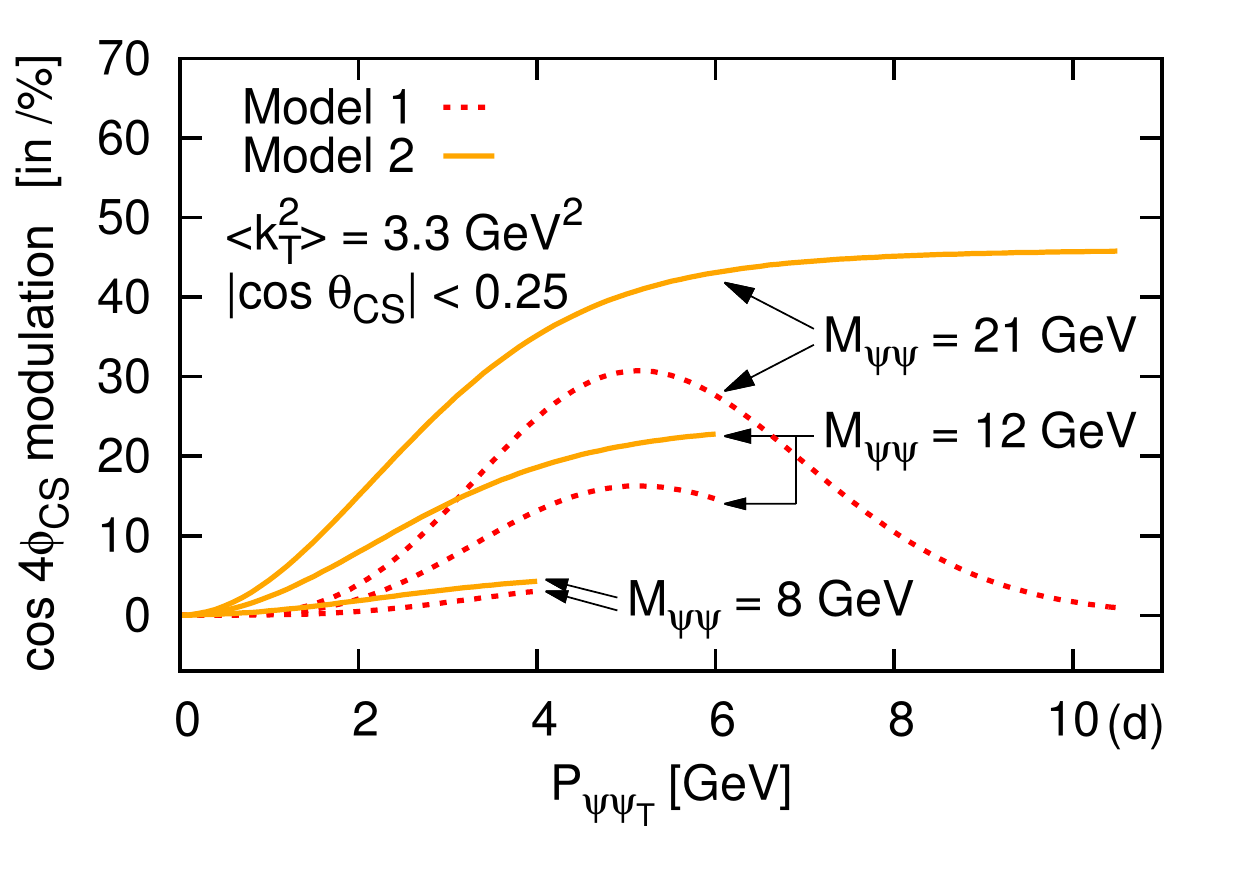}\label{fig:S4_costheta}}\vspace*{-.5cm}\\
\subfloat{\includegraphics[height=0.44\textwidth,angle=-90]{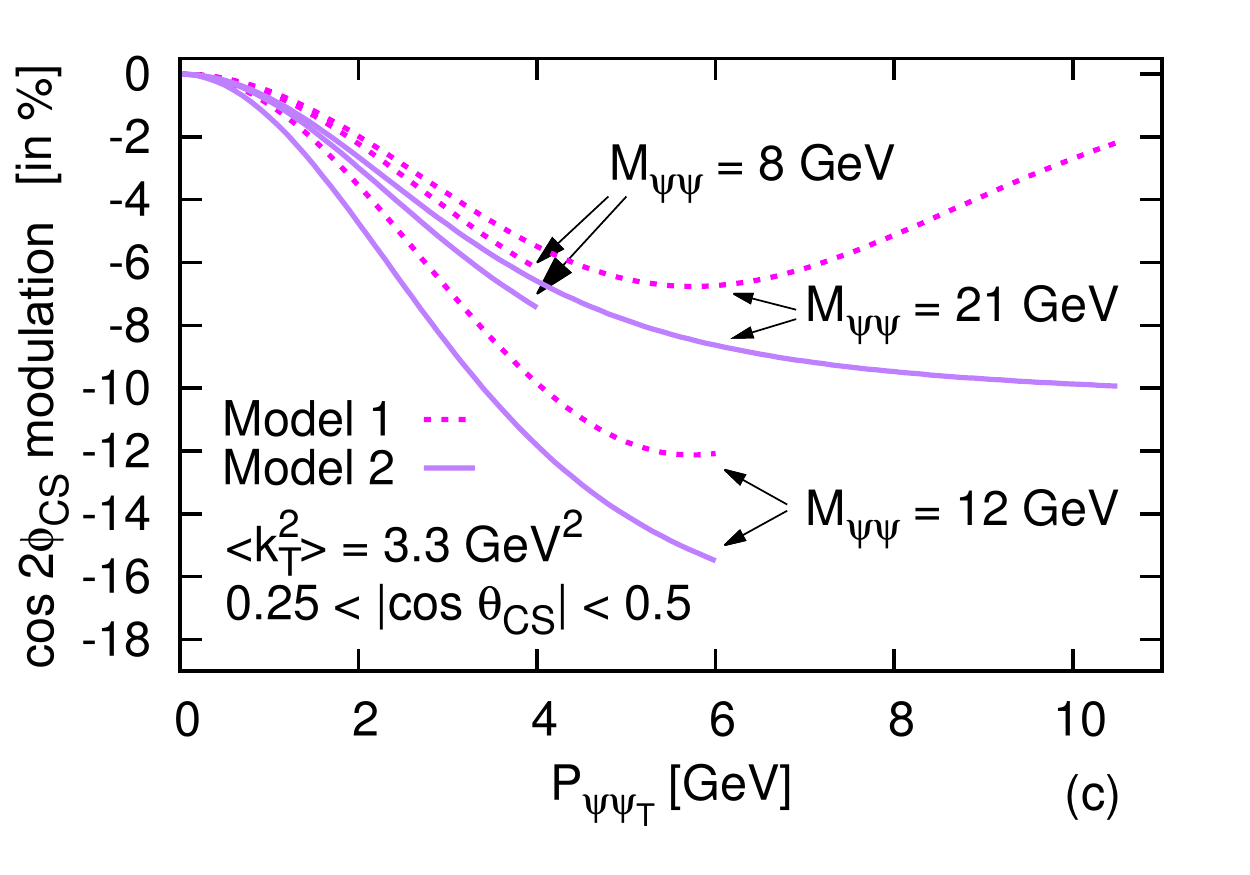}\label{fig:S2_costheta-2}}
\subfloat{\includegraphics[height=0.44\textwidth,angle=-90]{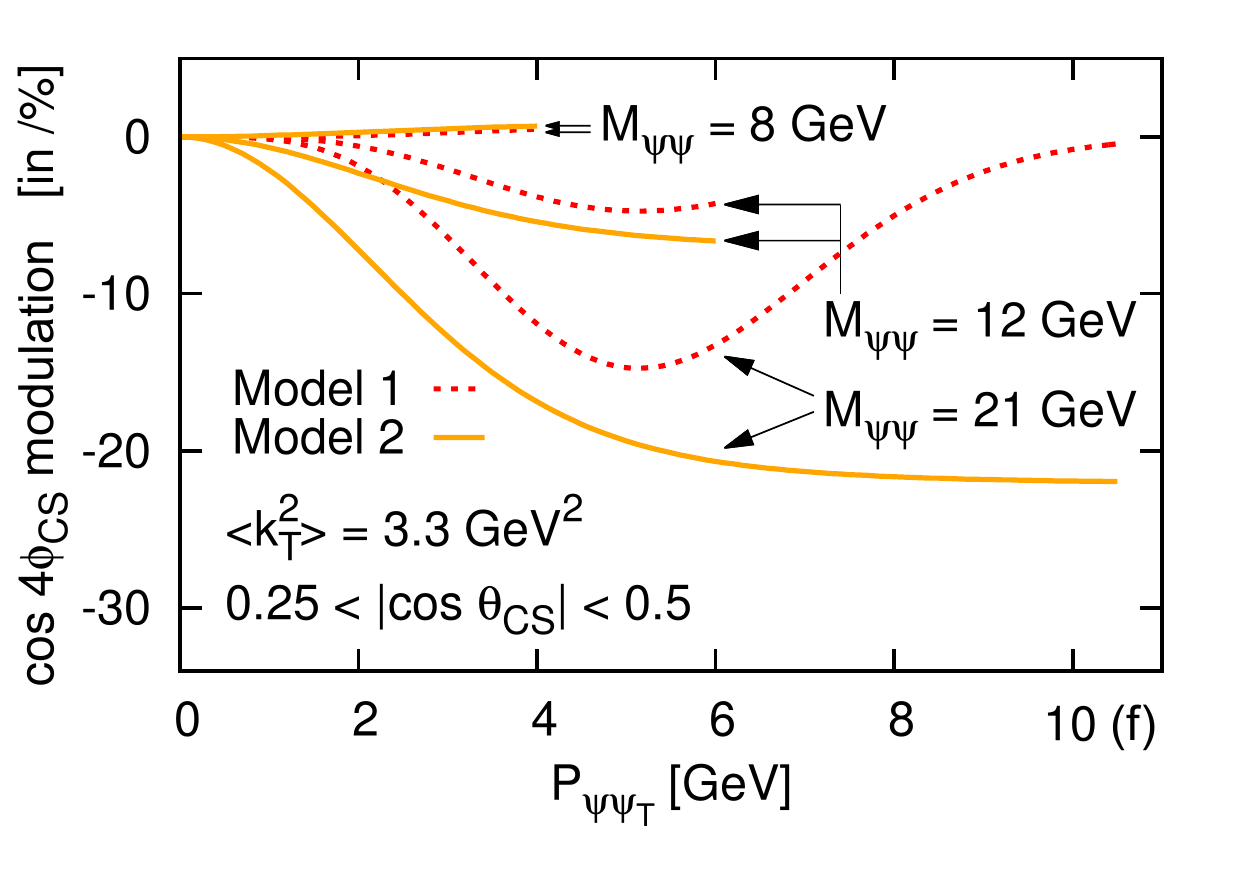}\label{fig:S4_costheta-2}}\vspace*{-.1cm}
\caption{$\cos n\phi_{\CS}$ modulations for $n=2,4$ computed for $|\cos\theta_{\CS}| < 0.25$ and for $0.25 < \cos \theta_{\rm CS}< 0.5$ for both our models of $h_{1}^{\perp g}$, for 3 values of $M_{\Q\Q}$ (8, 12 and 21 GeV) relevant respectively for the LHCb~\cite{Aaij:2016bqq}, CMS~\cite{Khachatryan:2014iia} and ATLAS~\cite{Aaboud:2016fzt} kinematics. The spectra are plotted up to $M_{\Q\Q}/2$. Our results do not depend on $Y_{\Q\Q}$.\label{fig:Si}}
\end{figure}

For $M_f$ close to $2M_{\Q}$ --its minimum value, $F_2 \to 3/{787} \times F_1$ and $  F_{3,4} \to 0$. 
Even though $F_2$ is not strictly zero, like in the $\Q+\gamma$ case, it is always very small and generates negligible
modulations to the $\qT$-differential cross section. Based on this, we performed in Ref.~\cite{Lansberg:2017dzg} 
the first extraction of $f_1^{g}$ using the latest LHCb data~\cite{Aaij:2016bqq}. Assuming a Gaussian $\kT$ dependence
we obtained $\langle \kT^2 \rangle = 3.3 \pm 0.8$~GeV$^2$ which encapsulates both non-perturbative and perturbative
effects since the scale relevant for such data sample is on the order of 8~GeV. 
Using this $\langle \kT^2 \rangle$ value and modellings of $h_1^{\perp g}$ such as the Gaussian form of Ref.~\cite{Boer:2011kf} (Model 1) or saturating the  positivity bound~\cite{Mulders:2000sh,Cotogno:2017puy} (Model 2), we obtained the $\cos n \phi_{\CS}$ modulations shown in \cf{fig:Si} where we note that the $\cos 2 \phi_{\CS}$ modulations are becoming larger for increasing $\cos\theta_{\CS}$.

\section{Conclusions}\vspace*{-0.2cm}

Quarkonium hadroproduction offers interesting possibilities to study gluon TMDs which only start to be investigated. We briefly reviewed here the cases of single-pseudoscalar-quarkonium production, vector-quarkonium pair production and associated production of a vector quarkonium with a photon or a $Z^0$ boson. The most promising case is that of $J/\psi+J/\psi$ which we expect to be studied soon at the LHC along the lines presented here, \ie\ with a dedicated measurement of the $\cos 2 \phi_{\CS}$ and $\cos 4 \phi_{\CS}$ modulations.

These would complement  --with a clean access to $f_1^{g}$ and $h_1^{\perp g}$-- future target-spin asymmetry studies (see \eg~\cite{Kikola:2017hnp,Hadjidakis:2018ifr})
to measure the gluon Sivers function $f_{1T}^{\perp g}$ as well as distributions of linearly-polarised gluons in a transversely polarised proton, $h_{1T}^{g}$ and $h_{1T}^{\perp g}$, allowing for a complete gluon tomography of the proton. 

\vspace*{-0.4cm}

\bibliographystyle{Science}

\bibliography{DIS2018-TMD}

\end{document}